\documentclass[letterpaper, 10 pt, conference]{ieeeconf}  

\IEEEoverridecommandlockouts                              

\usepackage{cite}
\usepackage{amsmath}
\usepackage{amssymb,amsfonts}
\usepackage{graphicx}
\usepackage{multicol}%
\usepackage{dblfloatfix}
\usepackage{grffile}
\usepackage{textcomp}
\usepackage{float}
\usepackage{booktabs}

 \usepackage[linesnumbered, ruled, vlined]{algorithm2e}
 \SetKwRepeat{Do}{do}{while}%
 \usepackage{bbm}

\usepackage{url}
\usepackage{color, colortbl}
\definecolor{Gray}{gray}{0.95}
\definecolor{LightCyan}{rgb}{0.88,1,1}

\DeclareMathOperator*{\argmin}{arg\,min}
\usepackage{bbm}
\usepackage{mathtools}
\usepackage{xcolor}
\usepackage{subcaption}

\usepackage{cleveref}
\crefformat{section}{\S#2#1#3}
\crefformat{subsection}{\S#2#1#3}
\crefformat{subsubsection}{\S#2#1#3}
\crefrangeformat{section}{\S\S#3#1#4 to~#5#2#6}
\crefmultiformat{section}{\S\S#2#1#3}{ and~#2#1#3}{, #2#1#3}{ and~#2#1#3}

\newtheorem{assumption}{Assumption}

\newcommand{\set}{\mathcal}

\DeclareMathOperator{\Tr}{Tr}

\let\oldref\ref
\renewcommand{\ref}[1]{(\oldref{#1})}

\SetKwComment{Comment}{$\triangleright$\ }{}
\SetCommentSty{itshape}
\title{\LARGE \bf
Data-Driven Control and Data-Poisoning attacks in Buildings:\\ the KTH Live-In Lab case study
}

\author{Alessio Russo$^{\star,1}$, Marco Molinari$^{2}$ and Alexandre Proutiere$^{1}$
\thanks{$^\star$ Corresponding author}
\thanks{$^{1}$Alessio Russo and Alexandre Proutiere are in the Division of Decision and Control Systems of the EECS School at KTH Royal Institute of Technology, Stockholm, Sweden.
        {\tt\small \{alessior,alepro\}@kth.se}}
\thanks{$^{2}$ Marco Molinari is in the Division  of Applied Thermodynamics and Refrigeration at KTH Royal Institute of Technology, Stockholm, Sweden.
        {\tt\small marcomo@kth.se}}
}

\begin{document}

\maketitle
\thispagestyle{empty}
\pagestyle{empty}


\begin{abstract}
This work investigates the feasibility of using input-output data-driven control techniques for building control and their susceptibility to data-poisoning techniques. The analysis is performed on a digital replica of the KTH Live-in Lab, a non-linear validated model representing  one of the  KTH Live-in Lab building testbeds. This work is motivated by recent trends showing a surge of interest in using data-based techniques to control cyber-physical systems. We also analyze the susceptibility of these controllers to data poisoning methods, a particular type of machine learning threat geared towards finding imperceptible attacks that can undermine the performance of the system under consideration. We consider the Virtual Reference Feedback Tuning (VRFT), a popular data-driven control technique, and show its performance on the KTH Live-In Lab digital replica. We then demonstrate how poisoning attacks can be crafted and illustrate the impact of such attacks. Numerical experiments reveal the feasibility of using data-driven control methods for finding efficient control laws. However, a subtle change in the datasets can significantly deteriorate the performance of VRFT.
\end{abstract}

\section{Introduction}
 Recent trends have shown a surge of interest in methods that \textit{intelligently learn from the data}. This trend is also motivated by recent successes in using deep-learning based methods for supervised learning tasks or control problems. In control systems data-driven control approaches, a branch of adaptive control, have gathered much attention over the last few decades \cite{campi2002virtual,hjalmarsson1998iterative,karimi2007non,de2019formulas,coulson2019data,esparza2011neural}, due to some interesting features, such as being able to directly compute a control law from experimental data gathered on the plant. This type of technique avoids identifying a model for the plant, which is particularly troublesome in those cases where it is difficult to derive, from first-principles, a mathematical description of the system, thus enabling direct data-to-controller design.
 
 In this work, we will analyze the feasibility of using the Virtual Reference Feedback Tuning (VRFT) method \cite{campi2002virtual,formentin2014comparison,esparza2011neural}  for temperature control in buildings. VRFT, compared to other data-driven control methods such as those based on Willems'  lemma \cite{willems2005note,de2019formulas}, allows to specify which requirements the closed-loop system should satisfy and aims at deriving a control law that satisfies the prescribed requirements. This particular feature of VRFT, coupled with the fact that the method is straightforward to use, makes it appealing in many control scenarios, from wastewater treatment \cite{rojas2012application} to unmanned aerial vehicle control \cite{invernizzi2016data} and control of solid oxide fuel cells \cite{li2011data}.
 
Despite these advantages, the performance of VRFT is tightly coupled with the data being used and can be seen as an identification problem. As such, it inherits the weaknesses of using data-based methods. For example, recently, it has been shown in the supervised learning community that a malicious agent can severely affect the performance of classifiers at test time by means of slight changes in the data used at training time \cite{biggio1,jagielski2018manipulating,goodfellow2014explaining}. A recent analysis demonstrated that data-driven control techniques are also affected by this particular attack for simple PID-like controllers \cite{alessio2020poisoning}, whilst the case where VRFT is used with non-linear controllers is left unexplored. Similar attacks, conducted at test time, have also been shown to work in the case of systems controlled through Reinforcement Learning controllers \cite{russo2019optimal}.\\

\textit{Contributions:}  the objectives of this work are twofold:\\
\textbf{(1)} We first analyze the feasibility of using VRFT for temperature control in buildings. This is validated by using a \textit{digital replica} of the KTH Live-In Lab testbed \cite{liveinlab}, a model of the real building set up using IDA Indoor Climate and Energy (IDA ICE) \cite{ida_ice}, a software used to simulate buildings performance.\\
\textbf{(2)} We then analyze the susceptibility of VRFT to data poisoning attacks, using the IDA ICE environment. 
We believe this is an important example of how data-driven control laws can be attacked. In buildings, the probability of sensors being hijacked is far from remote, and a malicious agent can use the data in several ways. This data could be used to determine the number of people present in the building or be poisoned to decrease the building's energy efficiency.   Gartner \cite{burke2019gartner}  predicts that through 2022 \textit{30\% of all AI cyberattacks will leverage training-data poisoning, AI model theft, or adversarial samples to attack AI-powered systems.} In \cite{kumar2020adversarial}, Microsoft engineers analyzed 28 companies and found out that only 3 of them have the right tools in place to secure their ML systems. This further stresses the importance of studying such problems.

\textit{Organization of the paper: }  \cref{sec:background} introduces the notation, the VRFT method, and the KTH Live-in Lab Testbed, which is a smart residential building located at the KTH campus. In \cref{sec:vrft_building}, the VRFT method is used to derive a controller that can control the temperature in the KTH Live-in Lab testbed's model. Finally, in \cref{sec:vrft_poisoned_building}, the data poisoning attack from \cite{alessio2020poisoning} is presented and applied to the VRFT method introduced in the previous section.

\section{Background and preliminaries}\label{sec:background}
\subsection{Notation}
We consider discrete-time models, indexed by $t\in \mathbb{N}_0$, and we will indicate by $[N]$ the sequence of integers from $0$ to $N$. We denote by $z$ the one-step forward shift operator and by $\mathcal{H}_2$ the Hardy space of complex functions which are analytic in $|z|< 1$ for $z\in \mathbb{C}$. For a vector $x\in \mathbb{R}^n$ and a function $f: \mathbb{R}^n\to \mathbb{R}$, we denote by $\nabla_x f(x)$ the $n$-dimensional vector of partial derivatives, where each element is $\partial_{x_i} f(x)$ with $\partial_{x_i} = \frac{\partial }{\partial x_i}$. We will describe a linear time-invariant system in the following way for $t\in \mathbb{N}_0$:
\begin{align}\label{eq:system_state_space1}
x_{t+1} &= Ax_t +Bu_t,\quad x_0\in \set X_0\\
y_t&=Cx_t+Du_t, \label{eq:system_state_space2}
\end{align}
where $x_t\in \mathbb{R}^n$ is the state of the system, $u_t\in \mathbb{R}^m$ is the exogenous input, $y_t\in \mathbb{R}^p$ is a vector of measurements and $\set X_0$ is a closed-convex subset of $ \mathbb{R}^n$. We can equivalently use transfer function notation and denote the input-output relationship using transfer function notation $y_t = G(z)u_t,$ with $G(z) = C(zI-A)^{-1}B+D$.  We also denote the multiplication of two transfer functions $G(z)$ and $L(z)$  by $GL(z)$ (similarly the sum). Finally, we will denote by $X_T = \begin{bmatrix}
x_0, \dots, x_T
\end{bmatrix}^\top$ a matrix of dimensions $(T+1)\times n$ containing a collection of state measurements of the system, for $T\in \mathbb{N}_0$. Similarly, we can define $U_T$ and $Y_T$.

\subsection{Virtual Reference Feedback Tuning}
In the following, we will denote by $\set D_N = (U_{N}, Y_{N})$ the data available to the learner that comes from experiments on the plant, with $N>1$. This data will be used to learn the control law, and it is usually assumed to have been taken in open-loop conditions. In VRFT \cite{campi2002virtual}, the design requirements are encapsulated into a reference model $M_r(z)$  that captures the desired closed-loop behavior from $r_t$ to $y_t$, where $r_t \in \mathbb{R}^p$ is the reference signal. We assume that $M_r$ satisfies some realizability assumptions, such as being a proper stable transfer function.

In VRFT, we wish to find a controller $K_\theta(z)$, parametrized by $\theta\in \mathbb{R}^{n_k}$,  that minimizes the difference between the reference model and the closed-loop system in the $\set H_2$ norm sense. Define $\Delta_\theta(z) = M_r(z)-[(I+GK_\theta)^{-1}GK_\theta](z)$, then the criterion is usually casted as follows
\begin{align}\label{eq:criterion_J_MR}
J_{\text{MR}}(\theta) &= \left\|M_r(z)-[(I+GK_\theta)^{-1}GK_\theta](z)\right\|_2^2\\
&= \frac{1}{2\pi} \int_{-\pi}^{\pi} \Tr\left[\Delta_\theta(e^{j\omega}) \Delta_\theta^\top(e^{-j\omega})\right]\textrm{d}\omega.
\end{align}
One can immediately observe that $J_{\text{MR}}(\theta)$ is non-convex in $\theta$. To address this difficulty, the following assumption \cite{campi2002virtual,karimi2007non,formentin2014comparison} is often used: 
\begin{assumption}[\cite{campi2002virtual}]\label{assumption1}
	The sensitivity function $I-M_r(z)$ is close (in the $\set H_2$ norm sense) to the actual sensitivity function $(I+GK_{\hat{\theta}})^{-1}(z)$ in the minimizer $\hat{\theta}$ of \ref{eq:criterion_J_MR}.
\end{assumption}

This allows us to instead consider the following criterion
\begin{equation}\label{eq:criterion_J}
\bar J_{\text{MR}}(\theta) = \left\|M_r(z)-[(I-M_r)GK_\theta](z)\right\|_2^2.\\
\end{equation}
One can show that minimizing \ref{eq:criterion_J} can be cast as a problem that involves minimizing the difference between the input signal $u_t$ injected during the experiments and the control signal $  K_\theta(z) e_t$ computed using the \textit{virtual error signal}, $e_t$. The latter is defined as $ e_t =  r_t -y_t = (M_r^{-1}(z)-1)y_t$ where $ r_t$ is the \textit{virtual reference signal} computed using the reference model $M_r(z)$ as $ r_t = M_r^{-1}(z)y_t$. Unfortunately this minimization will lead to a biased estimate of the minimizer if the controller that leads the cost function to zero is not in the controller set. To address this problem, one can introduce a filter $L(z)$ that will pre-filter the data $\set D_N$. One can then define the objective criterion that is actually solved in the VRFT method:
\begin{equation}\label{eq:criterion_J_VR_N}J_{\text{VR}}(\theta, \set D_N) = \frac{1}{N+1} \sum_{t=0}^{N} \|u_t - K_\theta(z)  e_t\|_2^2,\end{equation}
and it can be proven \cite{campi2002virtual} that for stationary and ergodic signals $\{y_t\}$ and $\{u_t\}$, we  get the following asymptotic result: 
$\lim_{N\to\infty}J_{\text{VR}}(\theta, \set D_N) =J_{\text{VR}}(\theta)$, where
\begin{align*}J_{\text{VR}}(\theta) &=  \frac{1}{2\pi} \int_{-\pi}^{\pi} \Tr\left[\bar \Delta_\theta(e^{j\omega}) \Phi_{u}(\omega)\bar\Delta_\theta^\top(e^{-j\omega})\right]\textrm{d}\omega\\
\bar \Delta_\theta(z) &\coloneqq I-[K_\theta(I-M_r)G](z),
\end{align*}
with $\Phi_{u}$ being the power spectral density of $u_t$. Let $K^\star$ denote the minimizer over all possible transfer functions $K(z)$ of $\left\|M_r(z)-[(I-M_r)GK](z)\right\|_2^2$. If $ K^\star\in \{K_\theta(z): \theta \in \mathbb{R}^{n_k}\}$, then $K^\star$ is also the minimizer of \ref{eq:criterion_J_VR_N}. Otherwise, one can properly choose a filter $L(z)$ to filter the experimental data so that the minimizer of \ref{eq:criterion_J_VR_N} and \ref{eq:criterion_J} still  coincide (refer to \cite{campi2002virtual} for details). Here the control set is assumed to be be linearly parametrized in terms of a basis of transfer functions:
\begin{assumption}\label{assumption2vrft_linear_control}
	The control law $K$ is represented by an LTI system $K_\theta(z)$ that is linearly parametrized by $\theta \in \mathbb{R}^{n_k}$, and we will write $K_\theta(z) = \beta^\top(z)\theta$, with $\beta(z)$ being a vector of linear discrete-time transfer functions of dimension $n_k$.
\end{assumption}

Assumption \oldref{assumption2vrft_linear_control} includes different types of control law, such as PID, and can be relaxed to other types of models, including neural networks  \cite{sala2005extensions,esparza2011neural}.

\subsection{KTH Live-In Lab Testbed and IDA ICE}
\begin{figure}[!tbp]
  \centering
  \begin{minipage}[b]{0.49\columnwidth}
    \includegraphics[width=\textwidth]{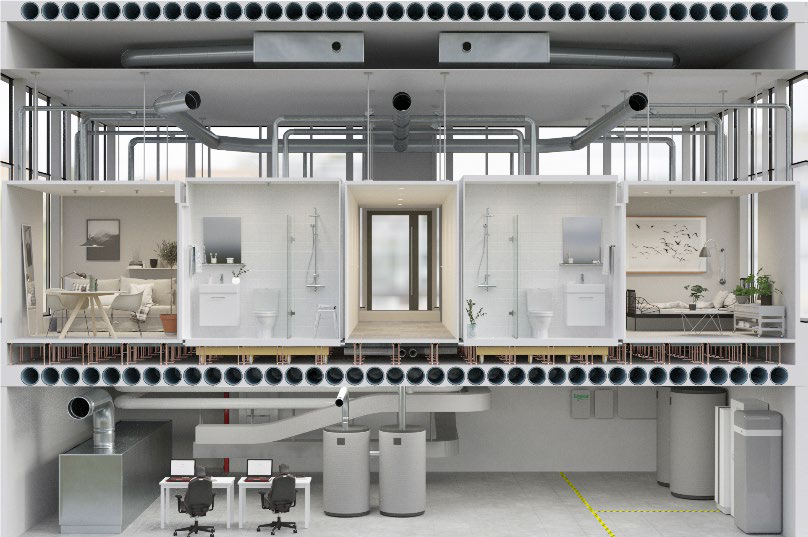}
    \caption{Digital view of the Live-In Lab Testbed KTH apartments.}
    \label{fig:kth_liveinlab}
  \end{minipage}
  \hfill
  \begin{minipage}[b]{0.49\columnwidth}
    \includegraphics[width=\textwidth]{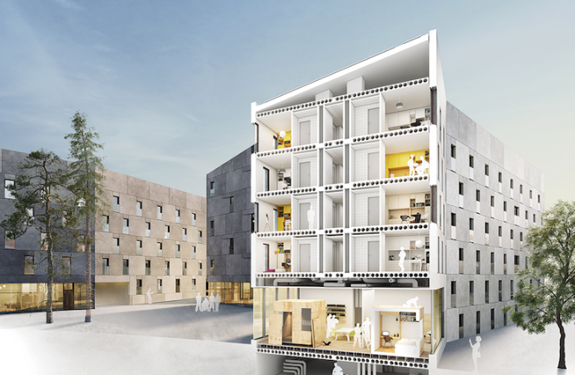}
    \caption{Digital image of the Live-In Lab [source: property developer Einar Mattsson].}
    \label{fig:liveinlab}
  \end{minipage}
\end{figure}

The Live-In Lab Testbed KTH \cite{liveinlab} (see Fig. \ref{fig:kth_liveinlab}) is located in one of Einar Mattsson's three plus-energy buildings (see Fig. \ref{fig:liveinlab}) in the KTH Main Campus, in Stockholm. The Testbed KTH premises feature a total of 305 m$^2$ distributed over approximately 120 m$^2$ of living space, 150 m$^2$ of technical space, and an office of approximately 20 m$^2$. The living space currently features four apartments; each apartment has a separate living room/bedroom and a bathroom and shares the kitchen as a common space. Space heating is provided via ventilation.
 The testbed, which is part of the larger Live-In Lab testbed platform, is designed to be energetically independent, with dedicated electricity generation systems through PV panels, heat generation (ground source heat pumps), and storage (electricity and heat) systems. Sensors are extensively used to monitor and control the indoor climate, to improve energy efficiency, study user behavior, and to improve control and fault detection strategies. 
 
 In this paper, a digital replica of the testbed that focuses on one apartment was created using the IDA ICE software \cite{ida_ice}. IDA ICE is a state-of-the art dynamic simulation software for energy and comfort in buildings. In order to assess the control laws that we derived, we set up a co-simulation environment that allowed IDA ICE and a Python script to communicate and exchange data through APIs available in IDA ICE.
\begin{figure}[b]
\centering
\includegraphics[width=\columnwidth]{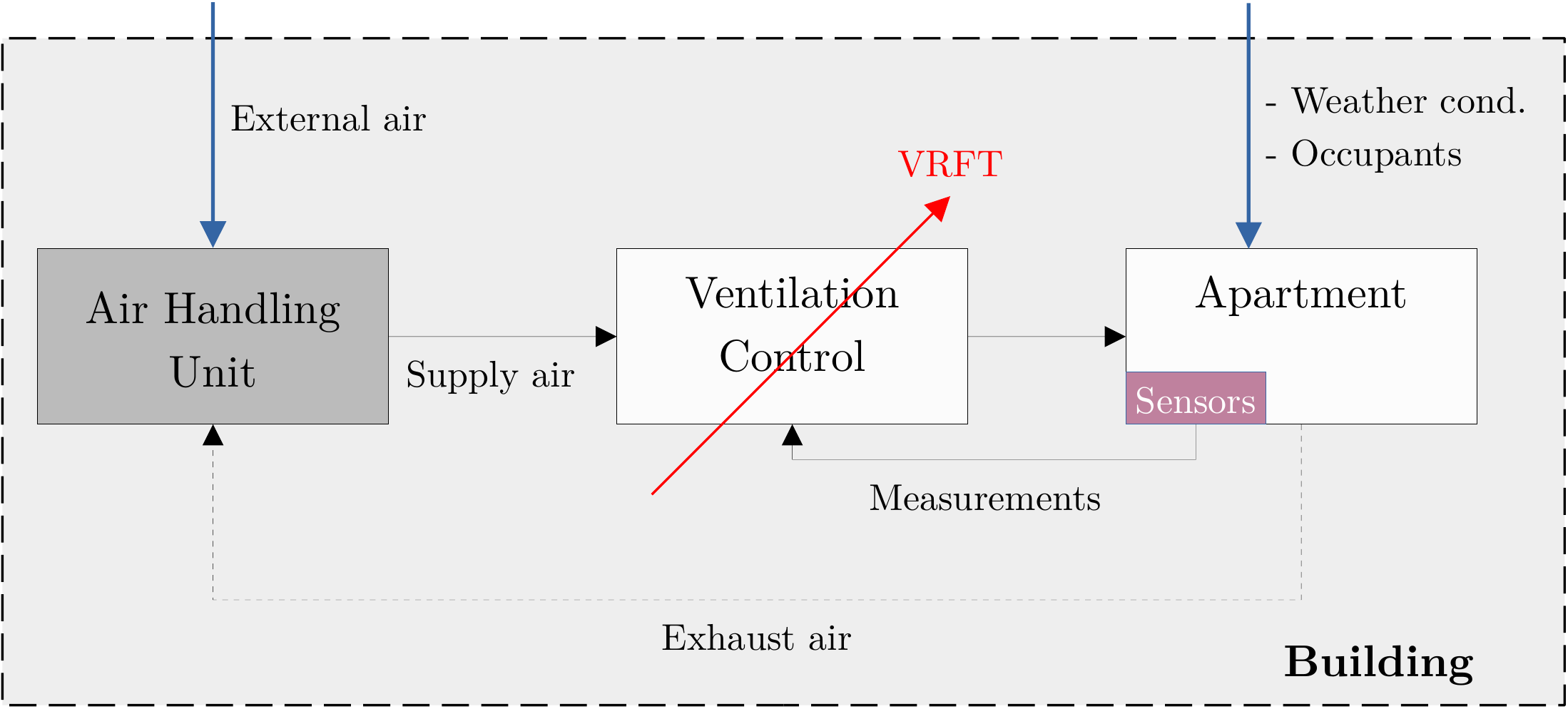}
\caption{HVAC architecture of the KTH Live-In Lab testbed.}
\label{fig:control_liveinlab}
\end{figure}

\begin{figure*}[b]
	\centering
  \includegraphics[width=\textwidth]{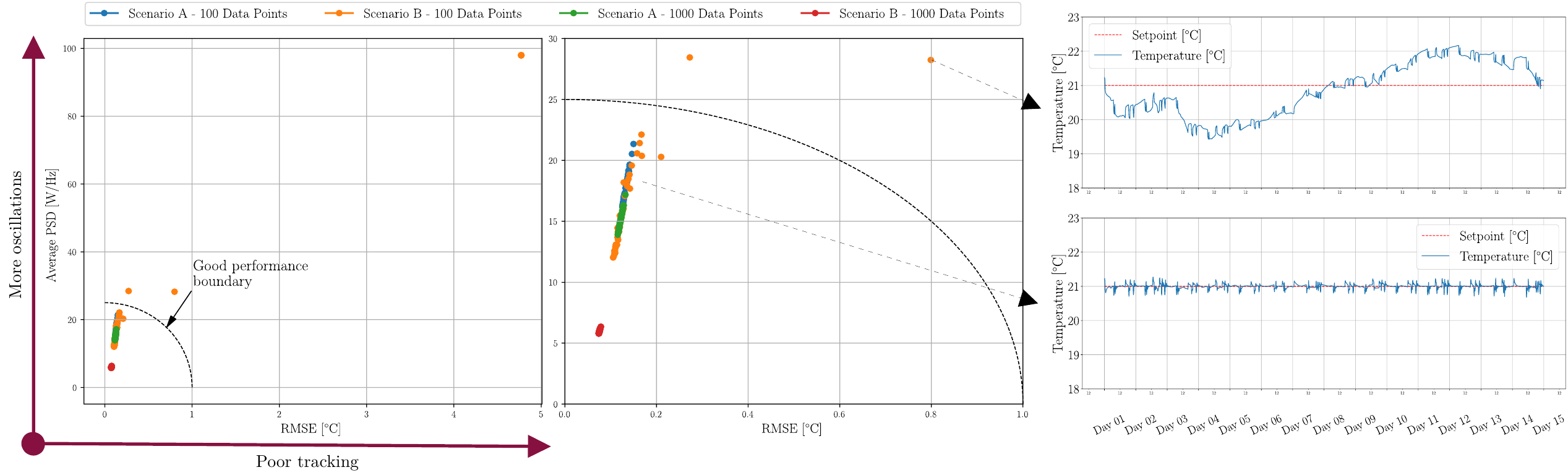}
  
  \caption{Results for the VRFT method. On the left are shown results distributed according to the RMSE and the average PSD computed over two weeks of data(for each color there are 50 simulations). Only scenario B with 100 data points yields poor performance controllers (6 of them have an RMSE that is roughly $5$ [$^\circ$C]). As an example, a controller from Scenario \textbf{B} is shown at the top right, whilst at the bottom right is shown a controller with good performance form Scenario \textbf{A}.}
  \label{fig:vrft_results}
\end{figure*}
\section{VRFT Method and Temperature Control} \label{sec:vrft_building}
In this section, we will briefly describe how the VRFT method has been applied to derive a controller. We will (1) sketch the HVAC (Heating, ventilation, and air conditioning) architecture of the testbed; (2) outline the usage  of VRFT; (3)  conclude with a performance analysis of the derived controllers.

\subsection{Method and experiments}

\textbf{HVAC architecture.} Fig. \ref{fig:control_liveinlab} shows a model of the HVAC architecture of the Live-In Lab Testbed KTH. VRFT will be applied to the ventilation control unit that regulates the amount of airflow supplied from the central Air Handling Unit (AHU) to the various apartments in the buildings. Measurements coming from the apartment include the temperature $T(t)$ and CO$_2(t)$ readings, sampled every $540$ seconds (9 minutes).\\

\textbf{Experiment setup. } The first step involves designing an experiment that permits the user to gather informative data from the plant. The data will then be used to compute a control law using the VRFT method. We have decided to gather data from an empty apartment during winter months, and have used weather data from the local weather station in Bromma. For simplicity, we have chosen the experiments to be conducted in open-loop, with a  control signal distributed according to a Gaussian distribution $\mathcal{N}(\mu, \sigma^2)$. Since the amount of airflow can be expressed as a percentage, the control law $u_t$ is automatically clipped between $0$ and $1$.

Because of this saturation effect, one needs to pay extra attention while designing the experiment. To that aim, we have designed two scenarios: scenario (\textbf{A}) where the mean of the control law is $\mu=0.5$ and the standard deviation is $\sigma=1/6$; instead, in scenario (\textbf{B}) we have  $\mu=0.5$ and $\sigma=1$. Scenario (\textbf{B}) represents the case where the user does not take into consideration the saturation effect. In contrast, scenario (\textbf{A}) guarantees that with $99\%$ probability the control action will belong between $0$ and $1$ (at the cost of having a crest factor of $3$). The amount of data gathered for the training process is  another important factor. Therefore we have also decided to consider two cases: one where we use $N=100$ data points (roughly 10 hours of data with a sampling time of $540$ seconds), and $N=1000$ data points (that is 150 hours). Finally, due to the experiment's randomness, we have decided to generate $50$ sets of simulations for each scenario.\\
\begin{figure}[H]
\centering
\includegraphics[width=0.9\columnwidth]{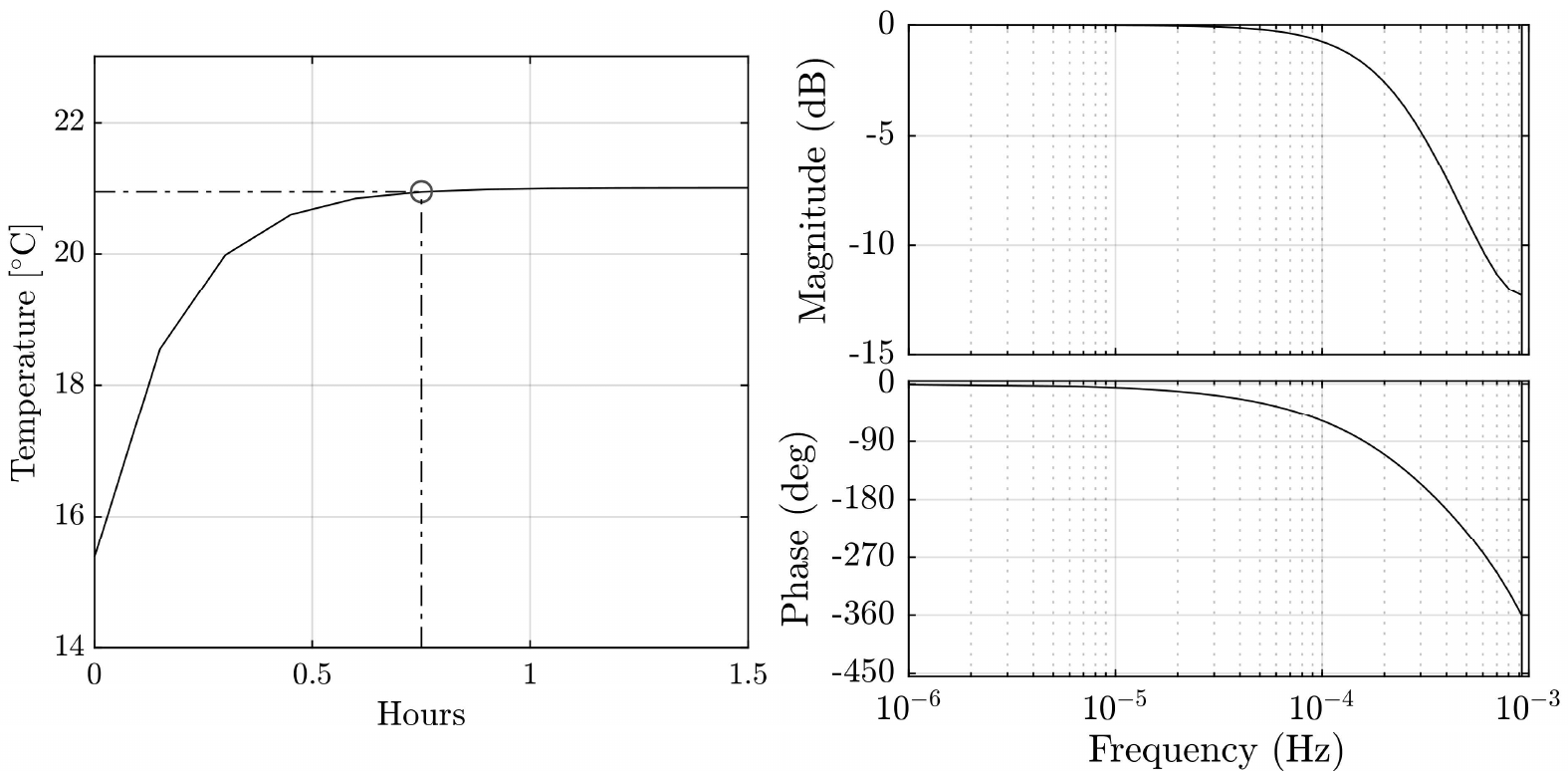}
\caption{On the left: step response of the reference model $M_r(z)$ (the circle denotes the settling time); On the right: Bode plot of the reference model $M_r(z)$.}
\label{fig:ref_model_plots}
\end{figure}

\textbf{Reference model and control law. } In VRFT, the user has to specify the closed-loop system's design requirements by choosing a specific reference model $M_r(z)$. This model, together with the data gathered during the experiments, is used to derive the control law $K_\theta(z)$. We have opted for a simple reference model and assumed that the closed-loop response of the system could be well represented by a second-order system.  In practice, we assumed that it would take approximately one hour for the heating system in consideration to increase the temperature in the apartment from $15^\circ$ to $21^\circ$ degrees Celsius. 

Therefore, we have chosen a reference model of the type \[M_r(z) = \frac{(1-\lambda)^2}{z^2-2\lambda z+\lambda ^2},\] where $\lambda = e^{-T_s\omega_0}$ with $\omega_0=0.002$ [rad/s] and $T_s=540 $ [s]. Fig. \ref{fig:ref_model_plots} shows the response of $M_r(z)$ to a step signal (with amplitude $21$, starting from an initial temperature of roughly $15^\circ$ [C]), and its Bode plot. All the data has been pre-filtered using a filter $L(z)=(1-M_r(z))M_r(z)$ (as explained in \cref{sec:background}; or see \cite{campi2002virtual} for more details). Finally, we have chosen to use a simple PID controller, of the form 
\[K_\theta(z) = \beta^\top(z)\theta = \sum_{k=1}^{3} \theta_i \underbrace{\dfrac{z^{-k+2}}{z-1}}_{\beta_k(z)}.\]
 This is one of the simplest controller that one can use with VRFT. Future work could also involve the analysis of more complex controllers, such as neural networks.


\subsection{Performance validation and results}
\begin{figure}[t]
\centering
\includegraphics[width=0.9\columnwidth]{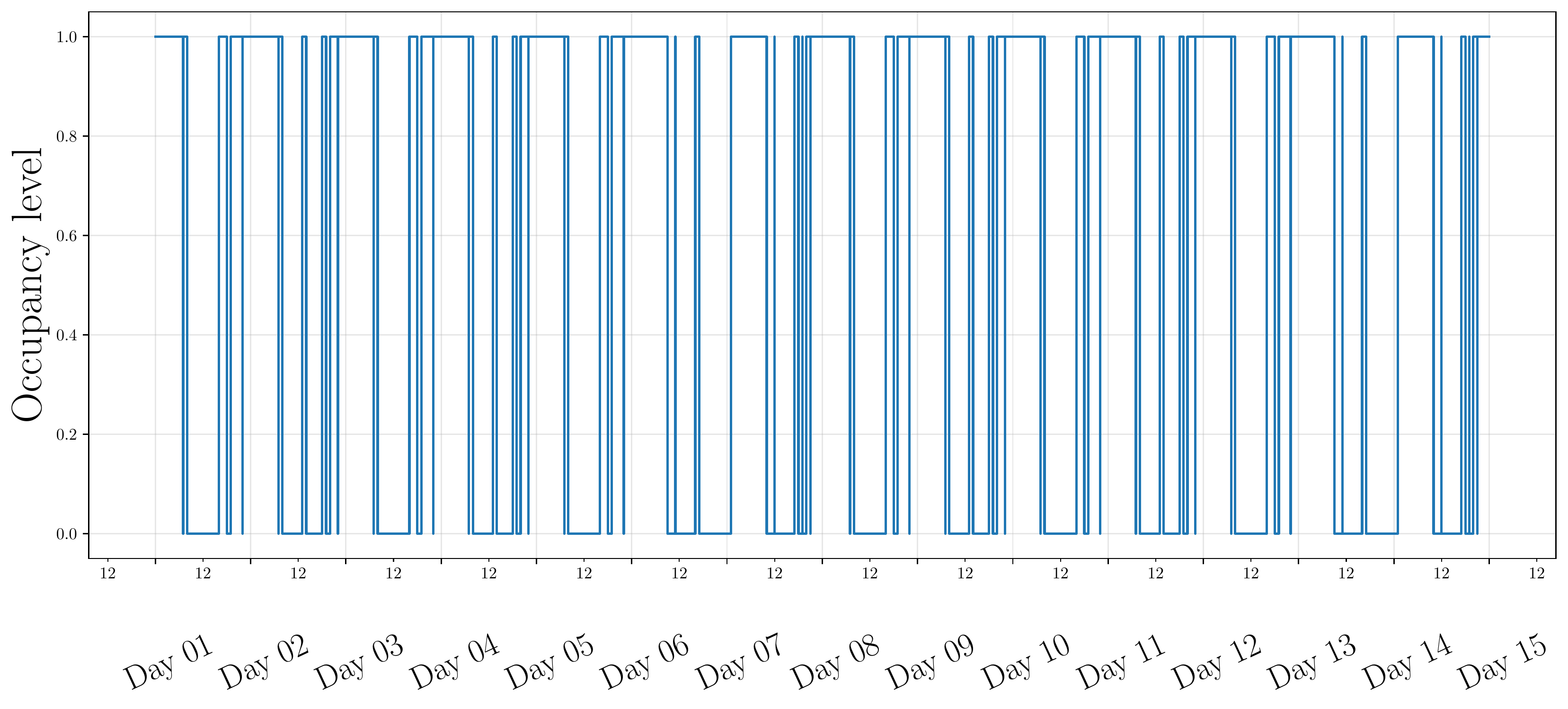}
\caption{Occupancy of the apartment over 2 weeks. We assumed there is only one person living in the apartment.}
\label{fig:occupancy}
\end{figure}

\textbf{Validation of the controllers. }As previously indicated, we have conducted 50 different simulations for each scenario, for a total of $200$ simulations. The performance of  each controller $K_{\theta_i}(z)$ has been validated over $2$ weeks (2240 data points), with the apartment being occupied by one person (according to the occupancy profile shown in Fig. \ref{fig:occupancy}).

\textbf{Performance criteria.} The performance of  a controller $K_{\theta_i}(z)$ has been evaluated on the basis of two criteria: (1) the RMSE of the temperature signal $e_{\text{RMSE}} = \sqrt{\frac{1}{N}\int_{0}^N (T(s) - r(s))^2 \textrm{d}s}$, where $T(t)$ is the temperature of the living room and $r(t)$ is the reference temperature, with constant value $r(t) =21$ [$^\circ$C]; (2) the average power spectral density of $T(t)$:  $e_{\text{PSD}} = \frac{1}{1/(2T_s)} \int_{0}^{1/(2T_s)} S_T(f) \textrm{d}f $, where $1/(2T_s)$ is the Nyquist frequency and $S_T(f)$ is the power spectral density (PSD) of the temperature $T(t)$ (which was computed using Welch's method).

\textbf{Results.} A summary of the results are shown in Fig. \ref{fig:vrft_results} and in Table \ref{table:vrft_results}. From visual inspection of the results, we decided to classify $K_{\theta_i}(z)$ to be a "good" controller if results for that controller satisfied the following ellipse condition $e_{\text{RMSE}}^2+(\frac{e_{\text{PSD}}}{15})^2\leq 1$: this guarantees that $K_{\theta_i}(z)$ satisfies good tracking performance and small oscillations. Overall, we found no major difference in performance in using 100 or 1000 data points for Scenario \textbf{A}, whilst there is a clear difference in using 100 or 1000 datapoints for scenario \textbf{B}. In the latter case, using fewer points may result in controllers with poor performance, as indicated by the results. Surprisingly, using $1000$ datapoints for scenario \textbf{B} results in high performance controllers. Nonetheless, the difference with controllers found in scenario \textbf{A} is minimal.

\begin{table}[H]
\centering
\resizebox{\columnwidth}{!}{%
\begin{tabular}{lcccc}
\hline
\rowcolor{Gray}\hline
\multicolumn{1}{l|}{\textbf{Scenario $(N=100)$}} & VRFT Loss   & RMSE  & Avg PSD & \%  good controllers     \\ \hline
\multicolumn{1}{l|}{Scenario A}                                                        &  $.36\pm .02$ & $.13\pm .01$    &  $17.97 \pm .31 $   &  $100\%$   \\
\multicolumn{1}{l|}{Scenario B}                                                     &   $.36\pm .02$  & $\mathbf{.70\pm .42}$   & $\mathbf{25.82\pm 7.45}$&  $84\%$   \\

\rowcolor{Gray}\hline
\multicolumn{1}{l|}{\textbf{Scenario $(N=1000)$}} & VRFT Loss   & RMSE  & Avg PSD & \%  good controllers     \\ \hline
\multicolumn{1}{l|}{Scenario A}                                                     &  $\mathbf{.38\pm .01}$  & $.12\pm .01$    &  $15.14\pm .18$&  $100\%$   \\
\multicolumn{1}{l|}{Scenario B}                                                     &  $.33\pm .01$  &$.08\pm .01$    &$5.95\pm .04$&  $100\%$   \\  \hline
\end{tabular}
}
\caption{For each scenario are shown the average value and confidence level at $95\%$ computed over $50$ simulations. Percentage of good controllers indicates the proportion of controllers  that falls inside the ellipse  $e_{\text{RMSE}}^2+(\frac{e_{\text{PSD}}}{15})^2=1$.}
\label{table:vrft_results}
\end{table}


\section{Data Poisoning of VRFT} \label{sec:vrft_poisoned_building}

In this section, we first present the data poisoning attack, introduced in \cite{alessio2020poisoning}, which inherits the main characteristics of the  attack formulated in \cite{biggio1}. We then apply the poisoning attack to the data that was gathered in the previous section and conclude with a performance analysis of the poisoned controllers.

\subsection{Attack Framework and setup}
\textbf{Attack formulation} We now assume that a malicious agent has access to the experimental data $\set D_N$ and knows the reference model $M_r(z)$ used by VRFT to identify a controller.
The goal of the malicious agent is to  degrade the performance of the resulting closed-loop system by subtly changing the dataset $\set D_N$.  

We denote the malicious signal on the actuators by $a_{u,t}\in \mathbb{R}^{m}$, and respectively by $a_{y,t}\in\mathbb{R}^{p}$, the attack signals on the sensors at time $t$.  The new input and output data points in the dataset at time $t$ are $u_t'=u_t+a_{u,t}$ and $y_t'=y_t+a_{y,t}$, respectively.  We will then denote the corrupted dataset by $\set D_N'=(U_N', Y_N')$  where $
U_{N}'=\begin{bmatrix}u_{0}' & u_{1}' &\dots & a_{N}'\end{bmatrix}^{\top}$, and similarly   $Y_N'=\begin{bmatrix}y_{0}' & y_{1}' &\dots & y_{N}'\end{bmatrix}^{\top}$. We will focus our attention on the maxmin attack, introduced in  \cite{alessio2020poisoning}, which is casted as a bi-level optimization problem
\begin{equation}\label{eq:op_maxmin}
\begin{aligned}
\max_{U_N',Y_N'}\quad &  J_{\text{VR}}(\hat \theta', \set D_N)& \\
\textrm{s.t.} \quad & \hat{\theta}' \in \argmin_{\theta} J_{\text{VR}}(\theta, \set D_N')\\
&\|U_N'-U_N\|_{2} \leq \delta_u,\quad \|Y_N'-Y_N\|_{2} \leq \delta_y,
\end{aligned}
\end{equation}
\begin{figure*}[t]
	\centering
  \includegraphics[width=\textwidth]{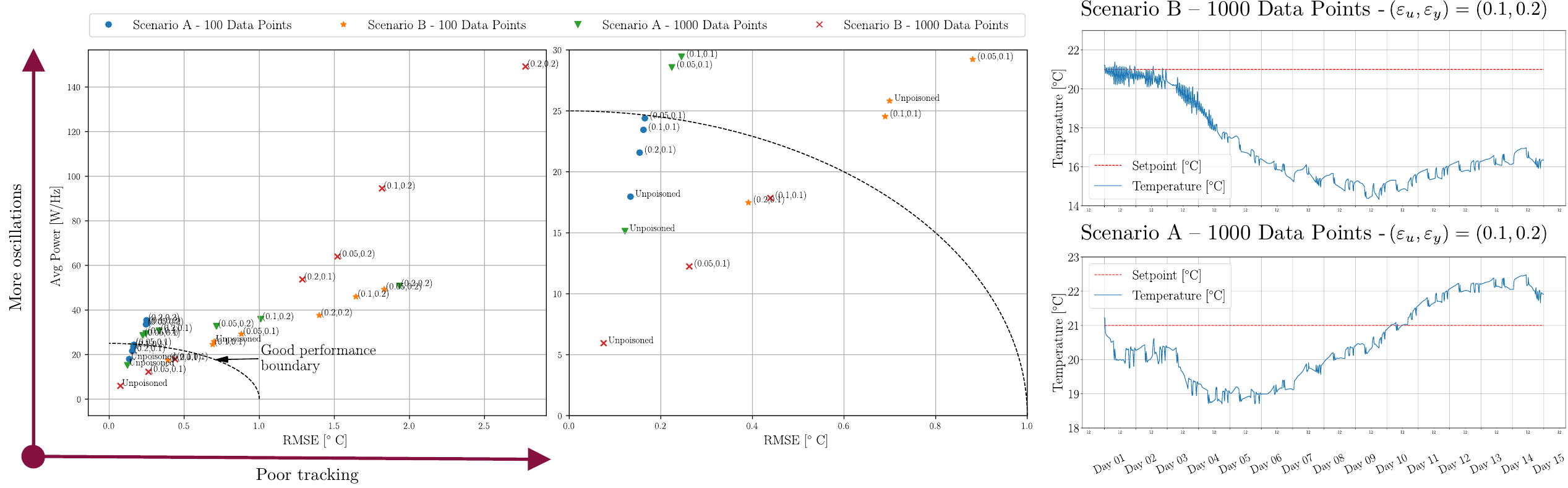}
  
  \caption{Results for the data poisoning attack. Each point on the left plots represents the average across 50 simulations for a specific set of values $(\varepsilon_u,\varepsilon_y)$, displayed on the top of each point (also the unpoisoned cases are depicted in the plots).}
  \label{fig:vrft_poisoned_results}
\end{figure*}
\noindent
where the constraints limit the amount of change applied to the dataset $\set D_N$.
In a maxmin attack, the malicious agent aims at maximizing the learner's loss. By choosing this cost function, the malicious agent is implicitly maximizing the residual error $\|M_r(z)-[(I-M_r)GK_\theta](z)\|_2$ (as $N\to \infty$). Despite this criterion's attractiveness, the resulting closed-loop system may remain stable or just slightly affected by the attack. One can formulate alternative criteria, as shown in \cite{alessio2020poisoning}, but for the sake of simplicity, we will restrict our analysis to the maxmin attack.

We also want to highlight a few differences compared to classical data poisoning: first, in contrast with supervised learning, there is no label for the data, which implies that we cannot merely maximize the probability of classification error. Second, the problem involves two sets of data, the input $U_N$, and the output data $Y_N$. Since the dependency of the solution may depend in a complicated way on $U_N$ and $Y_N$, the problem is harder.

\textbf{Convexity.} It can be shown that the optimization problem  \ref{eq:op_maxmin} is convex in $U'$ for a fixed $Y'$. Therefore, the maximum over $U'$, for some $Y'$, is attained on some extremal point of the feasible set. To find the optimal attack vector on the input, one can use, for example, disciplined convex-concave programming (DCCP) \cite{shen2016disciplined}. However, convexity with respect to $Y'$ does not hold, but one can still use gradient-based methods or genetic algorithms to find a solution.

\textbf{Algorithm and setup.} Based on the previous discussion, we use Alg. \ref{algo1} to approximately solve problem \ref{eq:op_maxmin}. We first perform the change of variable $A_u = U_N'-U_N$ and $A_y=Y_N'-Y_N$ and solve in the new variables $(A_u,A_y)$. The algorithm first solves \ref{eq:op_maxmin} in the input variable $A_u$ using DCCP, and then in the output variable $A_y$ using PGA (Projected Gradient Ascent). For both DCCP and PGA, we pick uniformly at random 20 initial points at every iteration. The algorithm stops whenever the increase between one iteration and the other is not greater than a fixed user-chosen value $\eta>0$.
\begin{algorithm}[h]
  \DontPrintSemicolon
  \KwIn{Dataset $\set D_N=(U_N, Y_N)$; parameters $\delta_u,\delta_y,\eta$}
  \KwOut{Poisoned dataset $\set D_N'$}
  $i \gets 0, (A_u^{(i)},A_y^{(i)})\gets (0,0)$ \Comment*[r]{Initialize algorithm}
  $\hat\theta^{(i)}\gets \argmin_\theta J_{\text{VR}}(\theta, U_N+A_u^{(i)},Y_N+A_y^{(i)})$\;
  \Do{$|J_{\text{VR}}(\hat\theta^{(i+1)}, U_N, Y_N)-J_{\text{VR}}(\hat\theta^{(i)}, U_N, Y_N)|>\eta$}{
		 $A_u^{(i+1)} \gets $ solve \ref{eq:op_maxmin} in $A_u$  using DCCP\;
		 $A_y^{(i+1)} \gets $ solve \ref{eq:op_maxmin} in $A_y$ (using $A_u^{(i+1)}$) with PGA\;

        $\hat\theta^{(i+1)}\gets \argmin_\theta J_{\text{VR}}(\theta, U_N+A_u^{(i+1)},Y_N+A_y^{(i+1)})$\;
        $i \gets i+1$\;
      }
  \Return $\set D_N'=(U_N+A_u^{(i)}, Y_N+A_y^{(i)})$
  \caption{Max-min attack algorithm}
  \label{algo1}
\end{algorithm}
\subsection{Performance and results.} 
\textbf{Setup. }
As in the previous section, we will consider 4 configurations: Scenario \textbf{A} with 100/1000 data points and similarly Scenario \textbf{B}. For each configuration, we chose $\delta_u$ and $\delta_y$ in \ref{eq:op_maxmin} as $\delta_u = \varepsilon_u \|U_N\|_2$ and $\delta_y = \varepsilon_y \|U_N\|_2 $, where $\varepsilon_u,\varepsilon_y$ are positive parameters in $[0,1]$ that we used as control knobs to vary the amount of change in $\set D_N$. For simplicity, we also assume that the data has already been pre-filtered using the filter $L(z)$.

\begin{figure}[H]
\centering
\includegraphics[width=0.94\columnwidth]{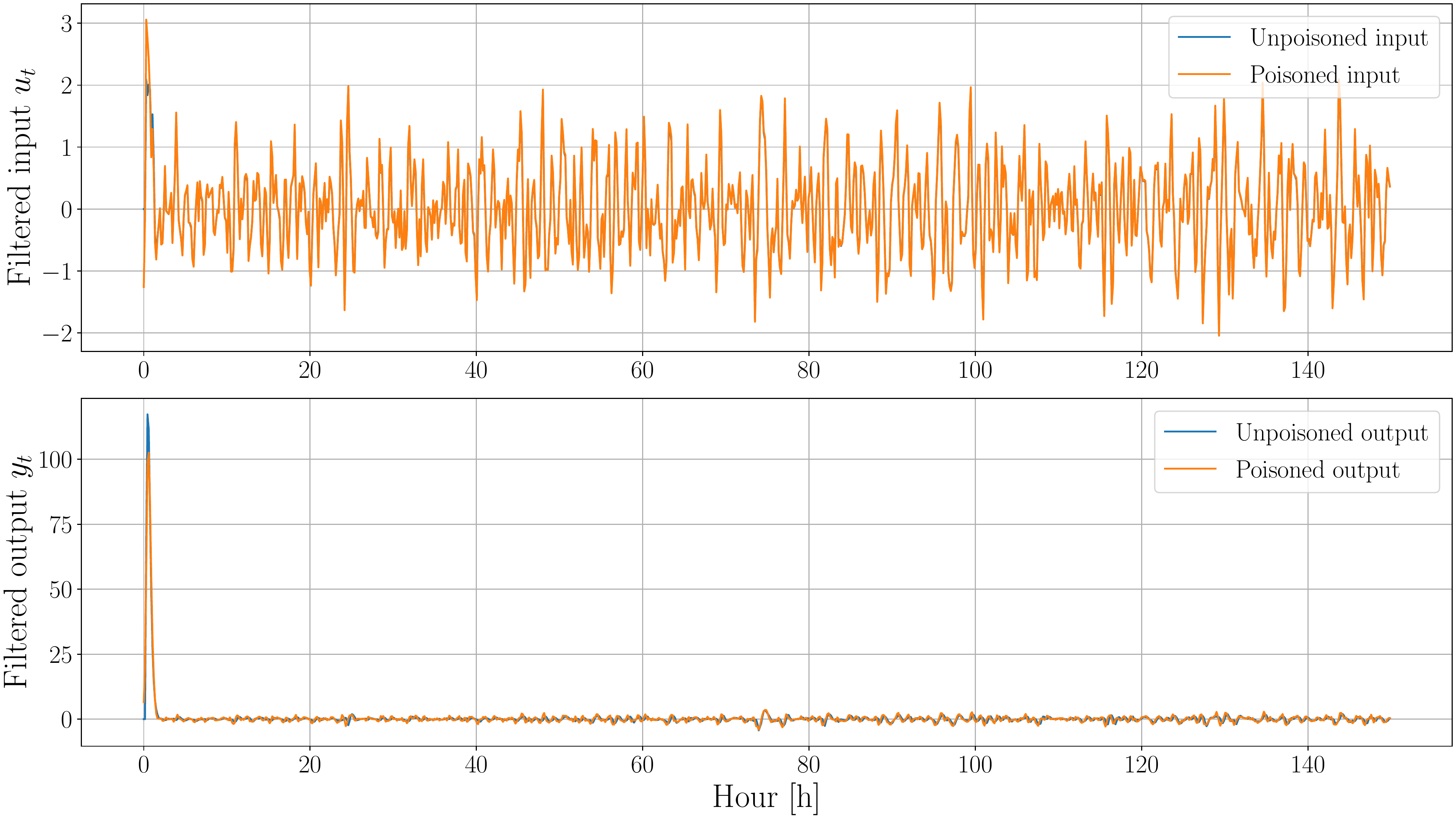}
\caption{Example of poisoned (pre-filtered) data for scenario \textbf{A} with 1000 data points and $(\varepsilon_u,\varepsilon_y)=(0.1,0.2)$.}
\label{fig:poisoned_data_example}
\end{figure}

\textbf{Results. } The main result is shown in Fig. \ref{fig:vrft_poisoned_results}, whilst in Fig. \ref{fig:poisoned_data_example} we show an example of poisoned dataset for Scenario \textbf{A}.

Due to the large number of simulations performed, we have decided to summarize results and show the average value for each configuration in Fig. \ref{fig:vrft_poisoned_results}. This means that each point in the left plots of Fig. \ref{fig:vrft_poisoned_results} represents the average across 50 simulations (on top of each point are written the values of $\varepsilon_u,\varepsilon_y$). The average values for the unpoisoned case are also shown, which can be used as reference values to understand the attack's impact. As expected, from the plot, one can immediately perceive that Scenario \textbf{B} is more susceptible to the attack. But Scenario \textbf{A}, for a large number of data points, is also significantly  affected by the attack, while using a low number of data points seems to improve robustness. Unfortunately, as depicted in Fig. \ref{fig:poisoned_data_example}, minimal changes lead to a substantial performance degradation, as shown in the bottom right plot in Fig. \ref{fig:vrft_poisoned_results}. This stresses the importance of performing experiments wisely and make sure that the gathered data is secured.
\section{Conclusion}
In this work, we have shown the feasibility of VRFT, an input-output data-driven method, for comfort control in buildings, namely temperature control, and analyzed the impact of the maxmin data poisoning attack. VRFT has been validated on a digital replica of the KTH Live-In Lab, modeled using IDA-ICE, showing good performance and small tracking error. We then analyzed the impact of data poisoning attacks, which revealed that small changes in the dataset could disrupt the controller's performance. Results also indicated that smaller datasets are more robust to data poisoning attacks, while datasets naively constructed are more susceptible to the attack, resulting in substantial performance degradation. This stresses the importance of securing the data used to derive the control law.

\bibliographystyle{IEEEtran}
\bibliography{IEEEabrv,ref}

\end{document}